\documentclass[prl,twocolumn,showpacs,preprintnumbers,amsmath,amssymb]{revtex4}
\usepackage{graphicx}
\usepackage{dcolumn}
\usepackage{bm}
\usepackage[tight]{subfigure}
\usepackage{enumerate}
\usepackage{amsmath}
\usepackage{verbatim}
\usepackage{color}

\begin{document}
\title{Coupling spin qubits via superconductors}
\author{Martin Leijnse}
\author{Karsten Flensberg}

\affiliation{
  Center for Quantum Devices, Niels Bohr Institute,
  University of Copenhagen,
  Universitetsparken 5, 2100 Copenhagen, Denmark 
}

\begin{abstract}
We show how superconductors can be used to couple, initialize, and read out spatially separated spin qubits.
When two single-electron quantum dots are tunnel coupled to the same superconductor, the singlet component
of the two-electron state partially leaks into the superconductor via crossed Andreev reflection.
This induces a gate-controlled singlet-triplet splitting which,
with an appropriate superconductor geometry, remains large for dot separations within the superconducting 
coherence length.
Furthermore, we show that when two double-dot singlet-triplet qubits are tunnel coupled to a 
superconductor with finite charging energy, crossed Andreev reflection 
enables a strong two-qubit coupling over distances much larger than the coherence length.
\end{abstract}
\maketitle
\emph{Introduction.}
Spin qubits defined in quantum dots~\cite{Loss98}  have emerged as a promising candidate for quantum information 
processing, and a number of recent experiments~\cite{Koppens05, Petta05, Koppens06, Nowack07} have demonstrated long 
coherence times and fast one- and two-qubit gates.
If a qubit is encoded in the spin of one electron, single-qubit rotations can be implemented with time-dependent 
magnetic fields~\cite{Koppens06}, or with electric fields utilizing the spin-orbit coupling~\cite{Nowack07}, and
two-qubit gates by gated control of the exchange interaction between different dots~\cite{Loss98}.
An alternative is to encode a single qubit in a two-dimensional subspace 
of the spin state of two electrons in a double dot~\cite{DiVincenzo00, Petta05} (for example the singlet and $T_0$ triplet). 
Then, the exchange interaction together with a static difference in the Zeeman splitting on the individual dots allow for 
arbitrary single-qubit rotations~\cite{Taylor05}.  
Readout is provided by spin-to-charge conversion, where one spin-qubit state is prohibited from tunneling out of the dot,
either by a Zeeman splitting~\cite{Elzerman04} (in a single-spin qubit) or by Pauli spin blockade~\cite{Ono02, Petta05} 
[in a singlet-triplet (ST) qubit].

However, exchange-based couplings are limited to qubits in direct proximity with each other.
Any large scale spin-based quantum computer architecture will have to rely on an alternative coupling mechanism 
of spatially separated spin qubits.
Recently, much progress has been in coupling superconducting qubits via a microwave 
field~\cite{Blais04, Sillanpaa07, Majer07, Schoelkopf08_rev}
(circuit quantum electrodynamics).
The same principles have been considered for spin qubits~\cite{Imamoglu99, Childress04, Burkard06, Jin12, Petersson12, Frey12}, 
but unfortunately the coupling between the electron spin and the cavity field is rather weak.

The Coulomb interaction provides a possible coupling of spin qubits which is reasonably long-ranged.
This requires coupling first the spin and charge degrees of freedom, e.g., via spin-orbit coupling~\cite{Flindt06, Trif07}. 
Alternatively, a double-dot ST qubit can be operated
in a regime of asymmetric level positions, where the singlet wave function has some weight with both electrons on the same 
dot [i.e., close to the $(1,1) \rightarrow (0,2)$ transition, where $(m,n)$ refers to a state with $m$ ($n$) electrons on 
dot $1$ ($2$)]. The different charge configurations of the singlet and triplet states then enable a direct capacitive 
coupling of two ST qubits~\cite{Taylor05, Stepanenko07, Weperen11, Shulman12}.
The range of the capacitive coupling might be increased by placing floating metallic gates between the quantum 
dots~\cite{Flensberg10c, Trifunovic12}.
Recent proposals also suggest using ferromagnets~\cite{Trifunovic13} or Majorana bound states~\cite{Leijnse12b} for two-qubit 
couplings.
However, achieving strong two-spin-qubit couplings remains challenging.

In this work, we introduce a proposal which uses a superconductor to couple spin qubits defined in spatially separated quantum dots. 
We show that crossed Andreev reflection~\cite{Hekking94, Byers95, Hartog96, Recher01} 
(CAR) allows for a large and long-ranged coupling. In the case of a one-dimensional superconductor, which can be realized in a semiconducting 
nanowire proximity-coupled to a bulk superconductor, we show that the coupling remains large for quantum dot separations up to the superconducting 
coherence length.
In addition, similar to Pauli spin blockade, CAR allows for qubit initialization and readout. 
Using a Cooper pair box (CPB), i.e., a floating superconductor with a finite charging energy, opens up even more possibilities.
In a double-dot ST qubit, CAR introduces 
a coupling between the spin-qubit state and the charge on the CPB. This can be exploited for strong (hundreds of MHz) capacitive 
two-qubit couplings of ST qubits separated by many $\mu$m (the only separation restriction being that the CPB to which both qubits 
couple should have a sizable charging energy).

\begin{figure}[t!]
  	\includegraphics[height=0.47\linewidth]{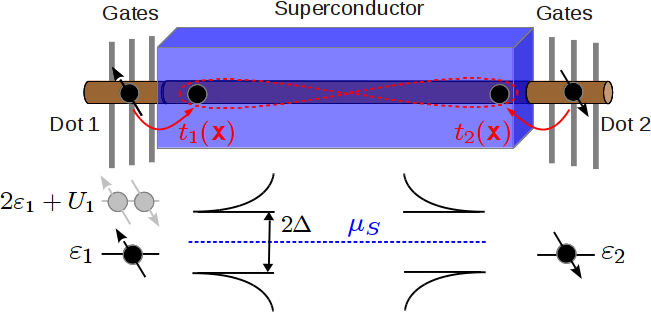}	
	\caption{(Color online) Coupling quantum dot spins via a superconductor.
	Upper panel: Two quantum dots defined within a semiconducting nanowire,  
	tunnel coupled to a superconductor (the main coupling is to the wire underneath the superconductor, which is 
	a one-dimensional superconductor because of the proximity effect).
	The grey gates control the dot energy levels and the tunnel amplitudes.
	A CAR process is illustrated, where one electron from each dot tunnels 
	into the superconductor and combines into a Cooper pair. 
	Lower panel: Energy-level sketch of the system.
	\label{fig:1}
 	}
\end{figure}

\emph{Two-dot coupling via a superconductor.} 
We start by considering two quantum dots tunnel coupled to different points of a superconductor, 
sketched in Fig.~\ref{fig:1}, similar to a so-called Cooper pair 
splitter~\cite{Recher01, Hofstetter09, Herrmann10, Hofstetter11, Das12b}.
In the sketch, the quantum dots are defined in a semiconducting nanowire (e.g., InSb or InAs), which is an experimentally promising system for spin 
qubits~\cite{Nadj10, Hu11}, and can form good interfaces with a superconductor~\cite{Doh05, vanDam06}. However, the results
are not restricted to this particular qubit realization.
The system is described by the Hamiltonian $H = \sum_{i = 1,2} ( H_{D i} + H_{T i}) + H_S$. Here, 
$H_{D i} = \sum_{\sigma} n_{i \sigma} \varepsilon_{i \sigma} + U_i n_{i \uparrow} n_{i \downarrow}$
describes quantum dot $i$, which has a single level with energy $\varepsilon_{i \sigma} = \varepsilon_i \pm B_i$ for spin projection 
$\sigma = \uparrow, \downarrow$,
occupation $n_{i \sigma} = d_{i \sigma}^\dagger d_{i \sigma}$, and Coulomb charging energy $U_i$. 
The BCS-type superconductor is described by 
$H_S = \sum_{\nu \sigma} E_{\nu \sigma} \gamma_{\nu \sigma}^\dagger \gamma_{\nu \sigma}$, where the Bogoliubov quasiparticle operators 
$\gamma_{\nu \sigma}$ are related to the electron operators $c_{\nu \sigma}$ and $c_{\nu \sigma}^\dagger$ in the standard way
and $E_{\nu\sigma} = \sqrt{\Delta^2 + (\varepsilon_{\nu \sigma}-\mu_S)^2}$, with $\mu_S$ being the chemical potential and 
$\Delta$ the superconducting gap. The coupling between dot $i$ and the superconductor is described by 
$H_{T i} = \sum_{\nu \sigma} \int d \mathbf{x} \; t_{i \nu \sigma} (\mathbf{x}) c_{\nu \sigma}^\dagger d_{i \sigma} + h.c.$,
where $\mathbf{x}$ is a coordinate in the superconductor.
For simplicity, spin-orbit interaction is not included in the model, but its qualitative effects will be discussed below.

We now want to calculate the superconductor-mediated coupling between electron spins on different dots,
cf., Ref.~\cite{Choi00} which discussed such coupling in a double-dot Josephson junction.
Focusing on the $(1,1)$ regime (each dot is occupied by a single electron), we should then calculate the energy shifts 
$\delta E_\alpha$ induced by the coupling to the superconductor, where $\alpha = S, T_0, T_{\pm}$ are the two-electron 
singlet and triplet states. 
We neglect all shifts which are the same for all $\alpha$, leaving fourth order terms in the tunnel Hamiltonian 
$H_T = \sum_i H_{T i}$ as the leading order contribution. We then find
\begin{eqnarray}\label{eq:Eshift} 
	\delta E_\alpha = \sum_n \frac{1}{E_\alpha - E_n} \left| \langle GS | \langle n | 
			  H_T \frac{1}{E_\alpha - H_0} H_T |\alpha\rangle |GS \rangle \right|^2, \nonumber \\
\end{eqnarray}
where $E_\alpha$ and $E_n$ are the unshifted energies of states $\alpha$ and $n$, $H_0 = \sum_i H_{D i} + H_S$,
and $|GS\rangle$ is the ground state of the superconductor (it can be verified that 
only intermediate states where the superconductor is in its ground state give an $\alpha$-dependent contribution).

There are two distinct contributions to $\delta E_\alpha$, which both lower the energy of the singlet relative 
to the triplet states: {\it (i)} Exchange, where an electron tunnels from one dot to the other via the superconductor and then back 
again, and {\it (ii)} CAR, where one electron from each dot tunnels into the superconductor to briefly combine into a Cooper pair 
before tunneling back again. These processes differ in the dot charge configuration in the intermediate state, 
where for exchange processes one dot is empty and the other doubly occupied ($|n\rangle = |02 \rangle$ or $|n\rangle = |20 \rangle$), 
while for CAR both dots are empty ($|n \rangle = | 00 \rangle$).
For simplicity we consider first zero magnetic field, $B_1 = B_2 = 0$.
For the CAR contribution, Eq.~(\ref{eq:Eshift}) gives
$\delta E_T^\mathrm{CAR} = 0$ for any of the triplet states $T = T_0, T_{\pm}$, while
$\delta E_S^\mathrm{CAR} = |\gamma|^2 / \varepsilon_\Sigma$, where $\varepsilon_\Sigma = \varepsilon_1 + \varepsilon_2 - \mu_S $. 
The CAR amplitude is given by
\begin{eqnarray}\label{eq:CARshift} 
	\gamma &=& \sum_{\nu i} \frac{\Delta}{\sqrt{2} E_\nu} \frac{1}{E_{\nu} - \varepsilon_{i}} 
			\int d \mathbf{x}_1 d \mathbf{x}_2 \; t_{1 \nu \uparrow}(\mathbf{x}_1) t_{2 \nu \downarrow}(\mathbf{x}_2). 
\end{eqnarray}
Assuming a finite tunnel coupling only between dot $i$ and a single point $\mathbf{x}_{i, 0}$ in the superconductor, 
we can write $t_{i \nu \sigma} = t_{i} \delta( \mathbf{x}_i - \mathbf{x}_{i, 0} )\psi_{\nu \sigma} (\mathbf{x}_{i, 0})$,
where $\psi_{\nu \sigma} (\mathbf{x})$ is a real space wavefunction in the superconductor. 
With the approximation $E_\nu - \varepsilon_{i} \approx E_\nu$ an analytic solution can easily be found 
(we come back to the full $\varepsilon$-dependent result below). 
For a three-dimensional ballistic superconductor~\cite{Falci01, Recher01}
one finds $\gamma \propto (k_F \delta x)^{-1} \mathrm{exp}(-\delta x / \pi \xi_0)$, 
where $\delta x = |\mathbf{x}_{1,0} - \mathbf{x}_{2,0}|$, $\xi_0$ is the superconducting coherence length, 
and $k_F$ is the Fermi wave vector of the superconductor. 
Although $\xi_0$ can be hundreds of nm, the algebraic prefactor limits the range of the interaction, 
since in typical metals $k_F^{-1} \sim$~\AA. The situation might be slightly better for a diffusive superconductor, where this prefactor is 
replaced by~\cite{Feinberg03} $(k_F \delta x)^{-1/2} (k_F l)^{-1/2}$, where $l$ is the mean-free path. However, the 
exponential suppression is more severe in this case, with $\xi_0$ being replaced by $\xi \approx \sqrt{\xi_0 l}$.

To achieve a long-ranged coupling one should reduce the dimensionality of the superconductor.
This makes the algebraic prefactor less severe, since it arises from summing over different paths  
between $\mathbf{x}_{1, 0}$ and $\mathbf{x}_{2, 0}$ with varying phases.
For a single-channel ballistic one-dimensional superconductor, we find [keeping $\varepsilon_{i}$ in 
Eq.~(\ref{eq:CARshift})]
\begin{eqnarray}\label{eq:CARshift1D} 
	\gamma &=&       \sqrt{2} t_1 t_2 \rho \sum_{i} \frac{\Delta}{\sqrt{\Delta^2 - \varepsilon_{i}^2}} 
	                 \left[ \pi + 2 \mathrm{tan}^{-1}\left( \frac{\varepsilon_{i}}{\sqrt{\Delta^2 - \varepsilon_{i}^2}} \right) 
			 \right] \nonumber \\
	       &\times&  \mathrm{sin} \left( k_F \delta x \right) e^{-\delta x / \pi \xi_0},
\end{eqnarray}
where $\rho$ is the normal-state density of states at the Fermi energy and we have assumed $\hbar^2 k_F^2 / 2m \gg \Delta$ and 
linearized the spectrum around $k_F$. 
There is no algebraic suppression and $\gamma$ remains large as long as $\delta x \lesssim \xi_0$.
In a setup as in Fig.~\ref{fig:1}, a one-dimensional superconductor results from proximity-inducing superconductivity in a 
semiconducting nanowire~\cite{Doh05, vanDam06}.
In fact, the experiments in Ref.~\cite{Hofstetter11} revealed an unexpectedly large CAR amplitude, likely due to the reduced 
dimensionality of the proximity-induced superconducting wire connecting the dots. 
\begin{figure}[t!]
  	\includegraphics[height=0.85\linewidth]{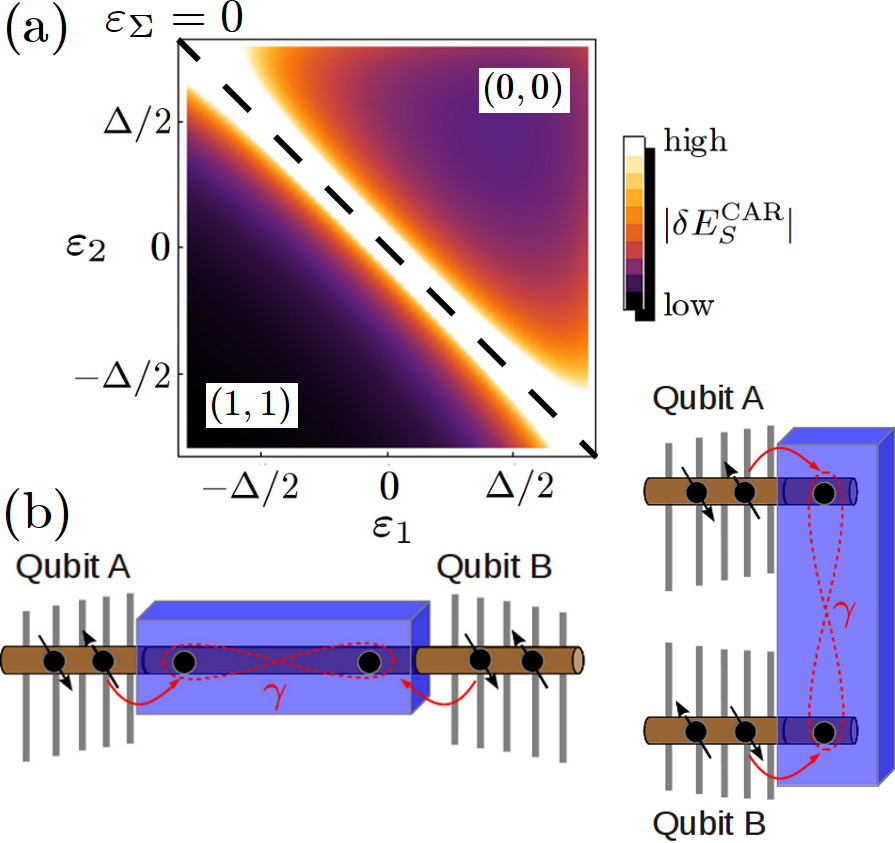}	
	\caption{(Color online) (a) $|\delta E^\mathrm{CAR}_S|$ as a function of $\varepsilon_1$ and $\varepsilon_2$,
	with $\gamma$ from Eq.~(\ref{eq:CARshift1D}), $\mu_S = 0$, and $B_1 = B_2 = 0$.
	At the CAR resonance (dashed line at $\varepsilon_\Sigma = 0$) the perturbatively calculated shift diverges.
	We focus on the regime  $\varepsilon_\Sigma < 0$, where the $(1, 1)$ charge configuration has lowest energy.
	(b) Two different systems with CAR-induced coupling between two ST qubits, defined either in the same wire, or in 
	different wires (the CAR amplitude is reduced in the latter case). \label{fig:2}
 	}
\end{figure}
Figure~\ref{fig:2}(a) shows the CAR-induced singlet-triplet splitting, $|\delta E^\mathrm{CAR}_S|$, as a function 
of $\varepsilon_1$ and $\varepsilon_2$, with $\gamma$ given by Eq.~(\ref{eq:CARshift1D}). 
The shift is of course largest close to the CAR resonance at $\varepsilon_\Sigma = 0$,
but we note that at a given $\varepsilon_\Sigma$, it can be increased by asymmetrically gating 
the dots, i.e., by operating at $\varepsilon_1 > \varepsilon_2$ or $\varepsilon_2 > \varepsilon_1$.
An analogous calculation of the exchange contribution gives $\delta E_T^\mathrm{ex} = 0$, 
$\delta E_S^\mathrm{ex} = \sum_i |\eta_i|^2 / (\varepsilon_\Sigma - E_{02,i})$, 
where $E_{02,i} = 2 \varepsilon_{i} + U$
and the exchange amplitude $\eta_i$ is given by an expression similar to Eq.~(\ref{eq:CARshift}).
$U$ is typically large, several meV in nanowire dots, and we neglect the exchange contribution in the following 
(it merely gives rise to an additional singlet-triplet splitting, which is small 
and lacks significant gate dependence around our operating point with large CAR). 

Finite magnetic field and/or spin-orbit interaction does not qualitatively change the above results. 
The situation is analogous to coupling between $(1,1)$ states and the $(0,2)$ singlet in double dots without 
coupling to a superconductor~\cite{Danon09}. At zero magnetic field, only one of the four $(1,1)$ states
can undergo CAR. 
At finite magnetic field, two $(1,1)$ states split off in energy similar to $T_{\pm}$ but in the presence of spin-orbit interaction
acquire a finite CAR amplitude, except for at specific field angles~\cite{Danon09, Nadj12}. 
CAR splits the remaining two degenerate states into one which can undergo CAR 
and one which cannot, and we can still refer to the former as $S$ and the latter as $T_0$.

In a setup as in Fig.~\ref{fig:1}, CAR can be used for controlled spin manipulation by for a certain time either moving 
the dot energy levels close to $\varepsilon_\Sigma = 0$, or always staying close to resonance but 
controlling the tunnel couplings $t_{1,2}$.  
However, to minimize decoherence due to charge noise, one should avoid moving too close to $\varepsilon_\Sigma = 0$, where the singlet state 
acquires a significant $(0,0)$ component.
Initialization can be done by moving both initially empty dot levels down through the CAR resonance adiabatically (with respect to $\gamma$), which 
fills the double dot with two electrons in a singlet state (Cooper pair splitting). 
If each dot is tunnel coupled to a normal electrode biased to drive an electron current into the superconductor, 
the triplet states act as blocking states since they do not allow for CAR~\cite{Eldridge10}. Similar to Pauli spin blockade, this can be used to 
prepare triplet states, and for readout through spin-to-charge conversion. 
Readout can also be done through charge detection by adiabatically moving both dot levels up through the CAR resonance, thereby 
selectively emptying the singlet state. 

The geometry discussed here can be used to couple two single-spin qubits, in which case the shift 
$\delta E_S$ allows for two-qubit gates, or to define a non-local ST qubit, in which case $\delta E_S$ together 
with a dot-dependent Zeeman splitting allows for universal single-qubit rotations. 
Alternatively, we can consider one of the geometries suggested in Fig.~\ref{fig:2}(b) with two different ST qubits, $A$ and $B$, 
where each double-dot ST qubit is internally coupled by normal exchange, while there is a long-distance coupling between one 
dot in each qubit via the superconductor.
This situation is analogous to two exchange-coupled ST qubits~\cite{Klinovaja12b}, but with a more long-ranged interaction.

\emph{Cooper pair box and two-qubit couplings.}
A single ST qubit coupled to a superconductor can be described by the effective Hamiltonian
\begin{eqnarray}\label{eq:HeffST}
	H_{ST} &=& \varepsilon_\Sigma |S\rangle \langle S | + 
			\varepsilon_\Sigma |T_0\rangle \langle T_0 | + \gamma |S \rangle \langle 00 | 
			+ h.c., 
\end{eqnarray}
where we have integrated out the quasiparticle degrees of freedom of the superconductor, 
leaving an effective CAR amplitude $\gamma(\varepsilon_1, \varepsilon_2)$, given by Eq.~(\ref{eq:CARshift1D}) 
for a one-dimensional superconductor.

\begin{figure}[t!]
  	\includegraphics[height=0.8\linewidth]{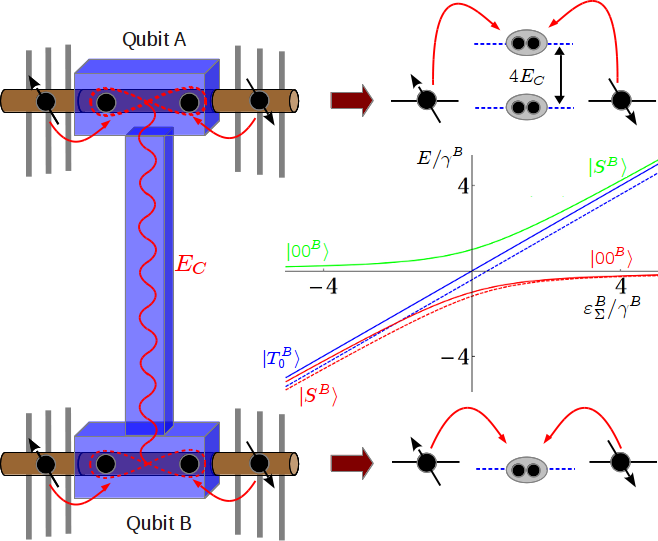}	
	\caption{(Color online) Proposed setup for two-qubit gates between two ST qubits.
	The energy-level sketch to the right shows the principle of the two qubit interaction:
	If the energy of qubit $B$ is adjusted so that the singlet component of the wavefunction (partially) leaks into the 
	CPB, through $E_C$ this affects the energy at which the singlet component of qubit $A$ can 
	do the same. Inset: Eigenspectrum of qubit $B$ as a function of $\varepsilon_\Sigma^B$ with qubit $A$ operated at 
	$\delta^A = -|\gamma^A| / \varepsilon_\Sigma^A = 1/2$, $E_C / 10 = \gamma^A = \gamma^B$, and neglecting the 
	gate dependence of $\gamma^B$. 
	The solid (dashed) lines represent the result when qubit $A$ is in state $|T_0^A\rangle$ ($|\tilde{S}^B\rangle$): 
	The difference in singlet-triplet splitting of qubit $B$ depending on the state of qubit $A$ allows two-qubit operations. 
	\label{fig:3}
 	}
\end{figure}

We now consider coupling instead to a small floating superconductor (CPB), with a well-defined number of Cooper pairs $N$ and 
a finite charging energy $E_C$. The Hamiltonian~(\ref{eq:HeffST}) remains valid in this case, but 
we need to also keep track of the superconducting charge state $|N\rangle$, and because charge is conserved in the total system, 
$|S/T_0\rangle$ should be interpreted as $|S/T_0\rangle |N_0 \rangle$, while $|00\rangle$ corresponds to 
$|00\rangle |N_0+1\rangle$. 
$\varepsilon_\Sigma$ is now measured relative to the energy of the $|N_0 \rangle \rightarrow |N_0 + 1\rangle$ transition.
Next, we consider two different ST qubits coupled to the same CPB, see Fig.~\ref{fig:3}, 
where the dots of the different qubits are separated by more than $\xi_0$, such that there is no CAR (or exchange) between them.
However, because of the conservation of total charge, a state where both double dots are empty must correspond to two 
Cooper pairs being added to the CPB. This gives rise to a capacitive coupling term 
\begin{eqnarray}\label{eq:2qubit}
	V^{AB} = 4 E_C |00^{A}\rangle |00^{B} \rangle \langle 00^{A}| \langle 00^{B}|, 
\end{eqnarray}
where superscripts refer to the two different ST qubits
(we will leave out the superscripts when referring to either one of the two qubits).
Equation~(\ref{eq:2qubit}) contains a projector unto the state where both double dots are empty, 
which is not a part of the normal computational space for an ST qubit. However, a qubit initially in 
$|S \rangle$ can be brought into $|00\rangle$ by adiabatically (with respect to $\gamma$) increasing $\varepsilon_\Sigma$
(moving along a red line in the spectrum in Fig.~\ref{fig:3}). If both qubits $A$ and $B$ are brought across the $|S \rangle \rightarrow |00\rangle$
avoided crossing for a finite time and then back again, Eq.~(\ref{eq:2qubit}) gives rise to a $\sigma_z^{A} \otimes \sigma_z^{B}$-type coupling.
The coupling strength is in principle only limited by the charging energy of the CPB, but
operating above the $|S \rangle \rightarrow |00\rangle$ anticrossing means that the spin superposition states 
have been fully converted into superpositions of charge degrees of freedom, resulting in short coherence times. 
It is likely better to stay below the anticrossing, where we can define effective computational states $|T_0 \rangle$ and 
$|\tilde{S}\rangle \approx (1/\sqrt{1+\delta^2}) |S\rangle + \delta |00 \rangle$, 
where $\delta = - |\gamma| / \varepsilon_\Sigma \ll 1$ (note that $\varepsilon_\Sigma < 0$).
Due to this small gate-dependent fraction of empty states which is added to the singlets, the two-qubit interaction to 
leading order in $\delta$ and $E_C / \varepsilon_\Sigma$ is given by 
\begin{eqnarray}\label{eq:2qubiteffective}
	V^{AB} &\approx& 4 E_C  (\delta^{A})^2 (\delta^{B})^2
		|\tilde{S}^{A}\rangle |\tilde{S}^{B} \rangle \langle \tilde{S}^{A}| \langle \tilde{S}^{B}| \\
		&\sim& 4 E_C  (\delta^{A})^2 (\delta^{B})^2 \sigma_z^A \otimes \sigma_z^B,
\end{eqnarray}
where the second line holds up to single-qubit rotations.
With $\gamma^{A} = \gamma^{B} = 50~\mu$eV~\cite{CARcomment}, and operating at 
$\delta^{A} = \delta^{B} = 1/3$, 
we find $V^{AB} \sim 0.5~\mu$eV$\sim 100$~MHz if $4 E_C = 50~\mu$eV.
This tiny value for $E_C$ would allow for a very long CPB.
Increasing $E_C$ increases $V^{AB}$, but 
Eq.~(\ref{eq:2qubiteffective}) only holds for $|E_C / \varepsilon_\Sigma| \ll 1$ and $V^{AB}$ saturates for larger 
$E_C$ [with the same $\gamma$ and $\delta$ as above, $V^{AB} \approx 300$~MHz for $4 E_C = 500~\mu$eV is found from Eqs.~(\ref{eq:HeffST}) and~(\ref{eq:2qubit})]. 
To reach large values while keeping $\delta$ small requires a large $\gamma$, which can be achieved in 
a few-channel nanowire with a  high-quality interface with the superconductor, as discussed above. 
Note that it is not necessary to have $E_C$ larger than temperature, as long 
as the CPB is well isolated from outside sources of quasiparticles, such that $N$ is conserved on the timescale of the two-qubit 
operations.

\emph{Conclusions.}
We have demonstrated that a superconductor can be used to mediate a long-distance coupling between spin qubits through CAR, which in a 
one-dimensional superconductor remains large for distances up to the superconducting coherence length.
Thus, a superconductor can be used to 
couple spatially separated spin qubits, or to define a non-local double-dot ST qubit. 
CAR-based initialization, manipulation, and readout would also provide definite evidence of the entanglement of 
electron spins in a Cooper pair splitter.
Furthermore, coupling two different ST qubits to the same CPB introduces a coupling between the spin qubits and the charge on the CPB,
which can be used for truly long-distance (many $\mu$m) two-spin-qubit gates with potential for fast operation 
(hundreds of MHz).

Finally, we mention that this charge--spin coupling has the potential to allow coupling of even more distant ST qubits through circuit 
quantum electrodynamics~\cite{Blais04, Majer07, Schoelkopf08_rev}. The two ST qubits in 
Fig.~\ref{fig:3} could be coupled to two different CPBs, which are then capacitively coupled to each other via a 
superconducting transmission line. This would allow the very large dipole moment of superconducting islands to be 
exploited to couple spins. 

\emph{Acknowledgements.} We thank C. M. Marcus, J. Danon, and A. Higginbotham for discussions.
The Center for Quantum Devices is funded by the Danish National Research Foundation.
\bibliographystyle{apsrev}

\begin{thebibliography}{48}
\bibitem[{CAR()}]{CARcomment}
\bibinfo{note}{With the estimate in Ref. [34] for the coupling to the
  superconductor, Eq. (3) gives $\gamma \approx 10 \;\mu$eV. However, the
  coupling between a nanowire and a superconductor can be made larger and may
  also be controlled by gate voltages, see for example Phys. Rev. Lett. \textbf{110}, 217005 (2013).}
\end{thebibliography}

\end{document}